\documentclass[%
aps,%
prb,%
twocolumn,%
showpacs]%
{revtex4}
\usepackage{graphicx}
\usepackage{bm}
\usepackage{color}%
\usepackage{amsmath,amssymb}

\newcommand{\tmfloatsmallb}[2]{
\begin{figure}[b]
#1
\caption{#2}
\end{figure}}

\newcommand{\tmfloatsmall}[2]{
\begin{figure}
#1
\caption{#2}
\end{figure}}

\newcommand{\tmtablesmall}[2]{
\begin{table}
#1
\caption{#2}
\end{table}}

\begin{document}

\title{Diffusion Monte Carlo: exponential scaling of computational cost for
large systems}

\author{Norbert Nemec}
\affiliation{Department of Physics,
 University of Cambridge,
 CB3 0HE, United Kingdom}

\date{\today}

\begin{abstract}
  The computational cost of a Monte Carlo algorithm can only be meaningfully
  discussed when taking into account the magnitude of the resulting
  statistical error. Aiming for a fixed error per particle, we study the
  scaling behavior of the diffusion Monte Carlo method for large quantum
  systems. We identify the correlation within the population of walkers as the
  dominant scaling factor for large systems. While this factor is negligible
  for small and medium sized systems that are typically studied, it ultimately
  shows exponential scaling. The scaling factor can be estimated
  straightforwardly for each specific system and we find that is typically
  only becomes relevant for systems containing more than several hundred
  atoms.
\end{abstract}


\maketitle

\section{Introduction}

Today's scientists can choose from a wide range of computational methods for
the simulation of quantum mechanical systems. These range from highly
efficient semi-empirial methods to density functional methods -- offering a
practical compromise of efficiency and accuracy -- all the way to very
accurate quantum chemical methods. Besides these deterministic methods,
various stochastic quantum Monte Carlo (QMC) methods are gaining ground,
offering exact handling of many strongly correlated systems and scaling up to
system sizes that are out of reach for the deterministic competitors.

The two major arguments that are typically brought up in the advocacy of QMC
are the excellent parallelizability and the good scaling behavior. Depending
on the QMC variant that is chosen, the collection of statistical data points
can be performed in parallel with little to no communication, making the
method well suited for high performance computers of all architectures. The
scaling behavior depends greatly on the details of the system and the method,
but it is generally found to be significantly better than that of quantum
chemical methods, and linear scaling algorithms have been
reported.\cite{williamson-lqmcc2001,manten-isidqmcwlmo2002,manten-lsftleiqmc2003,alf-lqmctwnlo2004,reboredo-onloflsqmcc2005,aspuru-guzik-asafteotleiqmc2005,kussmann-lfdqmcaftniiadms2008,nukala-afaeafsduiqmcs2009}

A commonly used method for the ab initio simulation of electronic structure is
diffusion Monte Carlo
(DMC),\cite{anderson-arsotseh31975,ceperley-gsotegbasm1980,foulkes-qmcsos2001}
typically using the fixed node approximation.\cite{reynolds-fqmcfm1982} For
this method, the bulk of the computational effort is spent on the repeated
evaluation of a trial wave function for electron positions that change step by
step, one electron at a time. The trial wave function is usually expressed as
a Slater determinant\cite{slater-ttocs1929} of single electron orbitals,
multiplied by a Jastrow factor\cite{jastrow-mpwsf1955} to express electron
correlations. For single electron orbitals expressed as maximally localized
Wannier functions,\cite{wannier-tsoeeliic1937} the local energy can be
reevaluated in constant time after a single electron move, leading to an $O
(N)$ algorithm for a complete time step of one
configuration.\cite{williamson-lqmcc2001,manten-isidqmcwlmo2002,manten-lsftleiqmc2003,alf-lqmctwnlo2004,reboredo-onloflsqmcc2005,aspuru-guzik-asafteotleiqmc2005,kussmann-lfdqmcaftniiadms2008,nukala-afaeafsduiqmcs2009}
As a further refinement to this, trial wave functions for DMC calculations are
today commonly expressed in a blip basis,\cite{alf-elbsfqmccocm2004} which
can be evaluated very efficiently.

In contrast to deterministic methods, however, the computational cost of a
Monte Carlo (MC) simulation is meaningless without specifying the statistical
error that is achieved. Deterministic methods typically have systematic errors
that are either intrinsic or depend on parameters that do not scale with the
system size. The statistical error of MC simulations on the other hand scales
very simply with the inverse square root of the run time while the scaling
with the system size is a nontrivial issue that depends on details of the
method and the system of study. Though the unfavorable scaling of the
statistical efficiency of DMC has been demonstrated
before,\cite{moroni-rmc2008} it has -- to our knowledge -- never been
studied systematically.

\tmfloatsmallb{\resizebox{\columnwidth}{!}{\includegraphics{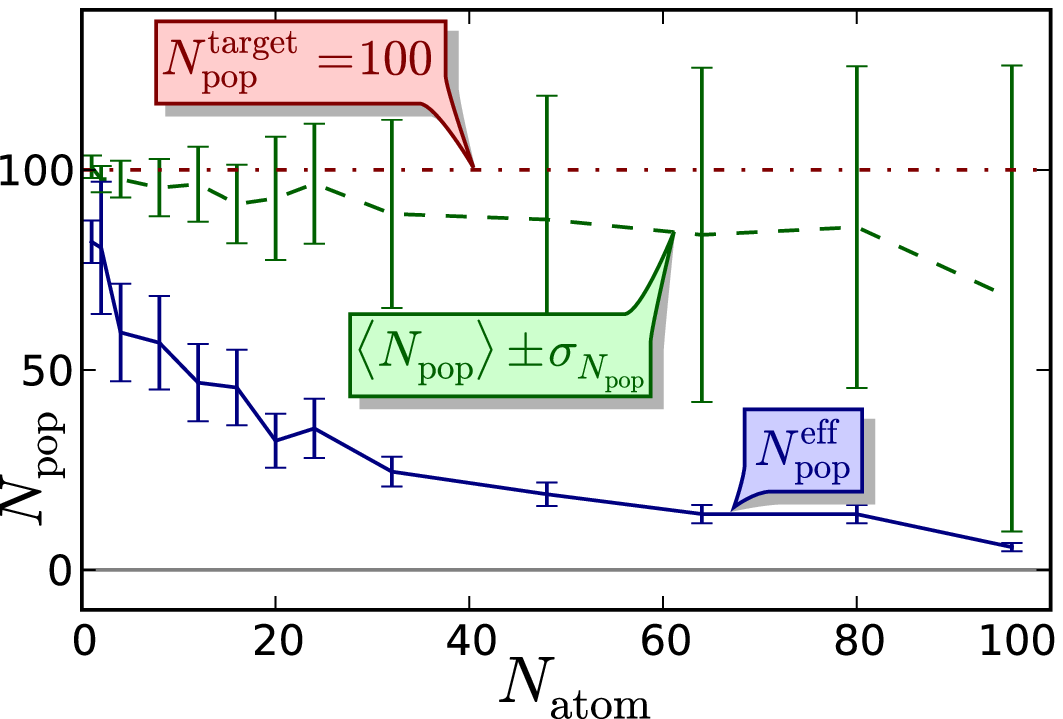}}}{(color
online) \label{fig:Npopeff}Scaling of the effective population size
$N_{\operatorname{pop}}^{\operatorname{eff}}$ [see Eq.~(\ref{eqn:Neffpop2})] in a sample
system [$\alpha = 1.5$, see Eq.~(\ref{eqn:detune})] with increasing number of
atoms $N_{\operatorname{atom}}$. The target population size is fixed, the true
population size $N_{\operatorname{pop}}$ fluctuates around a slightly lower average
(see text). The ``error bars'' of $N_{\operatorname{pop}}$ visualize the increasing
population fluctuations $\sigma_{N_{\operatorname{pop}}}$. The effective population
drops exponentially, due to increasing correlations within the population.}

In this paper, we will present a systematic study of the scaling behavior of
QMC calculations aiming for a fixed statistical error bar. The main focus will
be on the DMC algorithm including branching and population control as
described by Umrigar {\textit et~al.}~,\cite{umrigar-admcawvste1993} other
variants will be briefly discussed as well. The statements that we will derive
are expected to hold for DMC calculations in general, but to simplify
understanding, we will consider a ``typical'' system made up of $N$ similar
constituents which we simply call ``atoms''. This could be, for example, a
crystal in a simulation cell made up of $N$ primitive cells, a cluster of $N$
atoms or a large organic molecule of $N$ comparable groups.

We will begin by deriving several general quantities and continue by
demonstrating these in the case of a simple model of $N$ independent hydrogen
atoms. From this model we can numerically extract the missing pieces of the
scaling behavior, allowing a quantitative estimate of the scaling limit for
arbitrary systems. This limit will then be discussed for a number of different
sample systems.

\section{Scaling of computational cost}

The total computational cost of a DMC calculation (optionally split over a
number of parallel CPUs) is
\begin{eqnarray}
  t_{\operatorname{total}} & = & N_{\operatorname{step}} \times N_{\operatorname{pop}} \times
  t_{\operatorname{step}},  \label{eqn:ttotal1}
\end{eqnarray}
where $N_{\operatorname{step}}$ is the number of steps in imaginary time,
$N_{\operatorname{pop}}$ is the average population size and $t_{\operatorname{step}}$ is the
CPU time needed for one single all-electron step per configuration. Using a
so-called ``linear scaling'' QMC algorithm,\cite{williamson-lqmcc2001} each
all-electron move scales as
\begin{eqnarray*}
  t_{\operatorname{step}} & \propto & N,
\end{eqnarray*}
assuming that the evaluation of the Slater determinant dominates the
computational cost. For the moment, we assume that population fluctuations are
negligible and $N_{\operatorname{pop}}$ can be treated as an external parameter. The
influence of the population control will be discussed later on.

The standard error of the total energy can be expressed as
\begin{eqnarray}
  \delta E_{\operatorname{total}} & = & \sqrt{\left(
  \frac{\tau_{\operatorname{corr}}}{\tau_{\operatorname{step}}} \frac{1}{N_{\operatorname{step}}}
  \right) \left( \frac{\chi_{\operatorname{pop}}}{N_{\operatorname{pop}}} \right)
  \sigma_{\operatorname{dmc}}^2},  \label{eqn:Etotal}
\end{eqnarray}
with the constituents explained in the following:

The raw DMC variance $\sigma_{\operatorname{dmc}}^2$ is the variance of the local
energy of individual configurations over the whole simulation. Being based on
the mixed estimator, $\sigma_{\operatorname{dmc}}^2$ may deviate from the variance
$\sigma_{\operatorname{vmc}}$ of the trial wave function obtained in a VMC run. For
typical systems, however, we find that $\sigma_{\operatorname{dmc}}^2 \approx
\sigma_{\operatorname{vmc}}^2$. Assuming that the trial wave function of the whole
system can be optimized to about the same quality as that of a single
constituent $N$, this variance scales as
\begin{eqnarray}
  \sigma_{\operatorname{dmc}}^2 & \propto & N,  \label{eqn:vardmc}
\end{eqnarray}
since the local energy is dominated by the sum of $N$ independent atomic local
energies.

The (integrated) correlation time of a series of data points $x_i$ is given
by\cite{wolff-mcewle2004}
\begin{eqnarray}
  \tau_{\operatorname{corr}} & = & \tau_{\operatorname{step}} \left( 1 + 2 \sum_{j =
  1}^{\infty} \frac{\left\langle x_i x_{i + j} \right\rangle_i - \left\langle
  x_i \right\rangle_i^2}{\left\langle x_i^2 \right\rangle_i - \left\langle x_i
  \right\rangle_i^2} \right),  \label{eqn:tcorr}
\end{eqnarray}
where $\left\langle \cdot \right\rangle_i$ denotes the arithmetic mean over
the index $i$. $\tau_{\operatorname{corr}}$ in units of the time step
$\tau_{\operatorname{step}}$ describes the factor by which the number of steps
$N_{\operatorname{steps}}$ has to be divided to correct the error bar of the result
for serial correlation. To obtain $\tau_{\operatorname{corr}} / \tau_{\operatorname{step}}$,
as an alternative to computing Eq.~(\ref{eqn:tcorr}) directly, one can also
use the reblocking method.\cite{flyvbjerg-eeoaocd1989} Though the system
may have various correlation time scales, some of which depend on the system
size, we find that the integrated correlation time that is responsible for the
reduction in the resulting accuracy is dominated by the shortest correlation
times which depend on local properties, such as the kind of nuclei in the
system, but not on the size $N$.

The population correlation factor $\chi_{\operatorname{pop}} \geqslant 1$ captures the
inefficiency of the process due to population correlation and fluctuation. We
will treat this factor as an unknown quantity for the moment and discuss it in
detail afterwards.

Using Eqs.~(\ref{eqn:ttotal1}) and (\ref{eqn:Etotal}) along with the discussed
scaling laws, we can express the scaling of the total computational cost as
\begin{eqnarray}
  t_{\operatorname{total}} & \propto & \frac{\text{$\chi_{\operatorname{pop}}$}}{\delta
  E_{\operatorname{atom}}^2},  \label{eqn:ttotal2}
\end{eqnarray}
so we see that -- apart from the factor $\text{$\chi_{\operatorname{pop}}$}$ -- DMC is
in fact a constant scaling method if a fixed standard error per atom $\delta
E_{\operatorname{atom}} = \delta E_{\operatorname{total}} / N$ is required, as it is the case
for example in the study of long-ranged correlations in periodic systems. Of
course, memory limitations or implementation issues will limit the size of
computable systems. Within these limitations, however, the constant scaling
behavior is not so surprising, considering that for collecting statistical
data, it does not make any difference whether you simulate $N$ weakly
interacting systems in parallel or a single system $N$ times as long. Both
result in the same statistical error for the same computational cost.

\section{Population correlation}

For small enough systems, the factor $\chi_{\operatorname{pop}}$ is close to one,
which may be the reason why, to our knowledge, a systematical study has never
been attempted before. When scaling up the system size, however, population
correlation becomes important and we need a better understanding of its
scaling.

The DMC algorithm is based on a drift-diffusion process with branching and
killing of configurations due to fluctuations in the local energy. A freshly
branched pair of configurations is identical and thereby fully correlated. In
the following drift-diffusion process, it takes some time to decorrelate,
leading to a fluctuating amount of correlation within the population at any
time.

We consider a DMC run over $N_{\operatorname{step}}$ time steps $i$. We will first
consider a simplified model with constant population of $N_{\operatorname{pop}}$
configurations $p$, each having a local energy $E_p^i$. A generalization
including population fluctuations will follow in the section below.

An effective population size $N_{\operatorname{pop}}^{\operatorname{eff}}$ can be defined as
the number of configurations that would result in the same variance of the
average as the correlated population. For a long DMC run, the raw DMC variance
can be estimated from the averages over all configurations at all time steps
as
\begin{eqnarray}
  \sigma_{\operatorname{dmc}}^2 & = & \left\langle \left\langle \left( E_p^i \right)^2
  \right\rangle_p \right\rangle_i - \left\langle \left\langle E_p^i
  \right\rangle_p \right\rangle_i^2,  \label{eqn:sigma-dmc}
\end{eqnarray}
while the variance of the population average is defined as
\begin{eqnarray}
  \sigma_{\operatorname{pop}}^2 & = & \left\langle \left\langle E_p^i
  \right\rangle_p^2 \right\rangle_i - \left\langle \left\langle E_p^i
  \right\rangle_p \right\rangle_i^2,  \label{eqn:sigma-pop}
\end{eqnarray}
with the averages abbreviated as $\left\langle \cdot \right\rangle_p = \sum_{p
= 1}^{N_{\operatorname{pop}}} \cdot / N_{\operatorname{pop}}$ and $\left\langle \cdot
\right\rangle_i = \sum_{i = 1}^{N_{\operatorname{step}}} \cdot / N_{\operatorname{step}}$.

In the case of an uncorrelated population, we would find
$\sigma_{\operatorname{pop}}^2 = \sigma_{\operatorname{dmc}}^2 / N_{\operatorname{pop}}$, so we can
measure the amount of correlation by defining an effective population size as
the ratio
\begin{eqnarray}
  N_{\operatorname{pop}}^{\operatorname{eff}} & = & \sigma_{\operatorname{dmc}}^2 /
  \sigma_{\operatorname{pop}}^2,  \label{eqn:Neffpop2}
\end{eqnarray}
where $N_{\operatorname{pop}}^{\operatorname{eff}} \leqslant N_{\operatorname{pop}}$ with equality
only in the case of a completely uncorrelated population. By collecting the
necessary data during a DMC run, $N_{\operatorname{pop}}^{\operatorname{eff}}$ can be computed
at negligible cost. Fig.~\ref{fig:Npopeff} displays the scaling of the
effective population size with increasing system size for a sample system.

Note that the average population $N_{\operatorname{pop}}$ is typically slightly lower
than the target population $N_{\operatorname{pop}}^{\operatorname{target}}$, because the
population control implemented in CASINO uses the linear average $\left\langle
E_p^i \right\rangle_p$of the energy instead of the exponential average $\ln
\left\langle \exp \left( E_p^i \right) \right\rangle_p$ which determines the
actual growth of the population. Apart from reducing the population size, this
has no effect on the statistics or the result.

\section{Population fluctuations}

Keeping the population size completely fixed as we had assumed in the previous
model gives rise to a population control bias. To reduce this bias, the
population size $N_{\operatorname{pop}}^i$ is allowed to fluctuate and only weakly
controlled. The average over all time steps then needs to be weighted. In the
simple population control mechanism considered
here,\cite{umrigar-admcawvste1993} the weights are simply defined by the
population size $w_i = N_{\operatorname{pop}}^i$ for each time step $i$. The resulting
total energy average of a DMC run is then
\begin{eqnarray*}
  E_{\operatorname{tot}} & = & \frac{1}{\sum_i w_i} \sum_{i = 1}^{N_{\operatorname{step}}} w_i
  \left\langle E_p^i \right\rangle_p .
\end{eqnarray*}
In more sophisticated schemes, $w_i$ and $N_{\operatorname{pop}}^i$ may be decoupled.
To estimate the variance of this weighted average, we can split off the
correlation time into a factor and use the estimator of the variance of a
weighted average. Viewing the local energies $E_p^i$ as random variables and
the weights $w_i$ as constants given by a long DMC run, we can express this as
\begin{eqnarray}
  \operatorname{var} \left[ E_{\operatorname{tot}} \right] & = &
  \frac{\tau_{\operatorname{corr}}}{\tau_{\operatorname{step}}}  \frac{1}{\left( \sum_i w_i
  \right)^2} \sum_{i = 1}^{N_{\operatorname{step}}} w_i^2 \operatorname{var} \left[
  \left\langle E_p^i \right\rangle_p \right] .  \label{eqn:varEtot}
\end{eqnarray}
In the same interpretation of $E_p^i$ as random variables, we can replace the
averages over time steps in Eqs.~(\ref{eqn:sigma-dmc}) and
(\ref{eqn:sigma-pop}) by statistical expectation values $\left\langle \cdot
\right\rangle$ and write for each single time step
\begin{eqnarray*}
  \operatorname{var} \left[ \left\langle E_p^i \right\rangle_p \right] & = &
  \sigma_{\operatorname{dmc}}^2 - \left\langle \left\langle \left( E_p^i \right)^2
  \right\rangle_p - \left\langle E_p^i \right\rangle_p^2 \right\rangle .
\end{eqnarray*}
Substituting this into Eq.~(\ref{eqn:varEtot}) results in a sum over
expectation values, so the $\operatorname{var} \left[ E_{\operatorname{tot}} \right]$ itself
can be written as the expectation value of a single expression which can be
expressed as a product
\begin{eqnarray*}
  \operatorname{var} \left[ E_{\operatorname{tot}} \right] & = & \left\langle
  \frac{\tau_{\operatorname{corr}}}{\tau_{\operatorname{step}}} \times
  \frac{1}{N_{\operatorname{step}}^{\operatorname{eff}}} \times
  \frac{1}{N_{\operatorname{pop}}^{\operatorname{eff}}} \times \sigma_{\operatorname{dmc}}^2
  \right\rangle,
\end{eqnarray*}
with generalized expressions for the effective step number and population size
\begin{eqnarray*}
  \frac{1}{N_{\operatorname{step}}^{\operatorname{eff}}} & = & \frac{\sum_i \left( w_i^2
  \right)}{\left( \sum_i w_i \right)^2}\\
  \frac{1}{N_{\operatorname{pop}}^{\operatorname{eff}}} & = & 1 -
  \frac{1}{\sigma^2_{\operatorname{dmc}}} \frac{\sum_i w_i^2 \left( \left\langle
  \left( E_p^i \right)^2 \right\rangle_{\operatorname{pop}} - \left\langle E_p^i
  \right\rangle_{\operatorname{pop}}^2 \right)}{\sum_i w_i^2},
\end{eqnarray*}
where the quadratic appearance of the weights makes the effective population
size sensitive to population fluctuations as well.

To estimate the statistical efficiency of the DMC algorithm, it is most useful
to combine both quantities into the definition
\begin{eqnarray}
  \chi_{\operatorname{pop}} & = & \frac{N_{\operatorname{step}} \times
  N_{\operatorname{pop}}}{N_{\operatorname{step}}^{\operatorname{eff}} \times
  N_{\operatorname{pop}}^{\operatorname{eff}}} .  \label{eqn:xipop-def}
\end{eqnarray}
For a DMC run that is sufficiently long that the set of weights $w_i$ is a
good representation of the statistical distribution, the quantity
$\chi_{\operatorname{pop}}$ is an unbiased estimator. It corresponds exactly to the
quantity in Eq.~(\ref{eqn:Etotal}) and remains directly proportional to the
total CPU cost according to Eq.~(\ref{eqn:ttotal2}).

\section{Asymptotics of $\chi_{\operatorname{pop}}$}

The asymptotic behavior of $\chi_{\operatorname{pop}}$ for weak population correlation
can be derived by a few simple arguments. Assume, for a moment, the local
energy $E_{\operatorname{loc}}$ of configurations to be normally distributed as
\begin{eqnarray*}
  p \left( E_{\operatorname{loc}} \right) & = & \frac{1}{\sigma_{\operatorname{dmc}} \sqrt{2
  \pi}} \exp \left( - \frac{E_{\operatorname{loc}}^2}{2 \sigma_{\operatorname{dmc}}^2} \right)
  .
\end{eqnarray*}
A single configuration with $E_{\operatorname{loc}} < 0$ will branch at a rate of $-
E_{\operatorname{loc}}$. Integrated over the distribution of $E_{\operatorname{loc}}$, this
leads to a branching rate per configuration of
\begin{eqnarray*}
  \tau_{\operatorname{branch}}^{- 1} & = & \int_{- \infty}^0 \mathrm{d} E_{\operatorname{loc}} p
  \left( E_{\operatorname{loc}} \right)  \left( - E_{\operatorname{loc}} \right)\\
  & = & \frac{\sigma_{\operatorname{dmc}}}{\sqrt{2 \pi}} .
\end{eqnarray*}
It is, of course, known that the true distribution of the local energy is far
from normal.\cite{trail-hreiqmc2008} However, we can significantly relax
the previous assumption, considering that we essentially obtained the ratio
between standard deviation and mean absolute deviation which holds
approximately for a wide range of
distributions.\cite{gorard-ra9dtaotmd2005}

Starting from an initially uncorrelated population of size $N_{\operatorname{pop}}$,
the population after branching is $N_{\operatorname{pop}} + 1$ with two identical
configurations. The population mean is equivalent to that over $N_{\operatorname{pop}}
- 1$ correlations of single weight and one of double weight which has the
variance
\begin{eqnarray*}
  \left( \sigma_{\operatorname{pop}}^2 \right)' & = & \sigma_{\operatorname{dmc}}^2
  \frac{\left( N_{\operatorname{pop}} - 1 \right) \times 1^2 + 1 \times 2^2}{\left(
  \left( N_{\operatorname{pop}} - 1 \right) \times 1 + 1 \times 2 \right)^2} .
\end{eqnarray*}
For $N_{\operatorname{pop}} \gg 1$, the effective population size after the branching
is therefore $\left( N_{\operatorname{pop}}^{\operatorname{eff}} \right)' = N_{\operatorname{pop}} -
1$.

To keep the population stable, the branching and killing rates have to be
equal. For a weakly correlated population, it is trivial to see that a killing
event reduces the (effective) population by 1 as well.

After branching, the two copies evolving independently take an effective time
of $\tau_{\operatorname{corr}} / 2$ to decorrelate. (Since $\tau_{\operatorname{corr}}$
measures the amount of statistical data that is lost due to serial
correlation, which is exactly the quantity that we want to measure for
population correlation as well.)

For $\tau_{\operatorname{branch}} \gg \tau_{\operatorname{corr}}$, the effective population is
reduced by 1 at a rate of $2 N_{\operatorname{pop}} \tau_{\operatorname{branch}}^{- 1}$ and
restored to $N_{\operatorname{pop}}$ within $\tau_{\operatorname{corr}} / 2$. On average this
gives
\begin{eqnarray*}
  N_{\operatorname{pop}}^{\operatorname{eff}} & = & N_{\operatorname{pop}} - 2 N_{\operatorname{pop}}
  \tau_{\operatorname{branch}}^{- 1} \tau_{\operatorname{corr}} / 2
\end{eqnarray*}
or, since population fluctuations can be neglected,
\begin{eqnarray}
  \left( \chi_{\operatorname{pop}} \right)_{\tau_{\operatorname{corr}} \sigma_{\operatorname{dmc}}
  \rightarrow 0} & \rightarrow & 1 + \frac{\tau_{\operatorname{corr}}
  \sigma_{\operatorname{dmc}}}{\sqrt{2 \pi}} .  \label{eqn:xipop-perturbative}
\end{eqnarray}

\section{Model DMC process}

To study the dependence of $\chi_{\operatorname{pop}}$ on the system parameters beyond
the perturbative regime, we have implemented the full DMC algorithm on top of
a minimal model of a correlated diffusion process. Each configuration is
reduced to a single random variable with a simple exponential autocorrelation
so that $\sigma_{\operatorname{dmc}}$ and $\tau_{\operatorname{corr}}$ are free parameters.
The value of the variable is used directly as the local energy for the
branching process.

The population control mechanism described by Umrigar et al.~introduces an
additional parameter $\tau_{\operatorname{ceref}}$ with the dimension of time
(implemented in CASINO as parameter \texttt{cerefdmc}, used in updating the
reference energy $E_{\operatorname{ref}}$; $\tau_{\operatorname{ceref}} / \tau_{\operatorname{step}}$
corresponds to $g$ in the original
publication\cite{umrigar-admcawvste1993}). To avoid frequent population
instabilities for extreme parameter settings, we have restricted the
population size to a window around the target population and decoupled the
total weight from the total population size outside of this window, allowing
recovery from explosions or starvation without introducing any additional
bias. All the results presented below are in the regime where this mechanism
has no significant impact on the efficiency.

With each run, one obtains the factor $\chi_{\operatorname{pop}}$ as a function of the
parameters $\sigma_{\operatorname{dmc}}$, $\tau_{\operatorname{corr}}$, $\tau_{\operatorname{ceref}}$
and $\tau_{\operatorname{step}}$. The result must be dimensionless, reducing the
number of relevant parameters by one. Furthermore, one is interested in the
limit $\tau_{\operatorname{step}} \rightarrow 0$. We find that for $\left.
\tau_{\operatorname{step}} \lesssim 0.1 \times \min (\tau_{\operatorname{corr}},
\tau_{\operatorname{ceref}}, 1 / \sigma_{\operatorname{dmc}} \right)$, the inefficiency factor
$\chi_{\operatorname{pop}}$ becomes practically independent of $\tau_{\operatorname{step}}$ in
all cases. We can combine the remaining free parameters into
$\sigma_{\operatorname{dmc}} \tau_{\operatorname{corr}}$ and $\tau_{\operatorname{ceref}} /
\tau_{\operatorname{corr}}$, leading to the results displayed in
Fig.~\ref{fig:modeldmc}.

Most significantly, we find that Eq.~(\ref{eqn:xipop-perturbative}) is not
only confirmed in the perturbative limit, but its exponential continuation
gives a strict lower limit for the inefficiency factor
\begin{eqnarray}
  \chi_{\operatorname{pop}} & \geqslant & \exp \left( \sigma_{\operatorname{dmc}}
  \tau_{\operatorname{corr}} / \sqrt{2 \pi} \right)  \label{eqn:xipop-exponential}
\end{eqnarray}
where the deviation from this exponential depends on the ratio
$\tau_{\operatorname{ceref}} / \tau_{\operatorname{corr}}$.

\tmfloatsmall{\resizebox{\columnwidth}{!}{\includegraphics{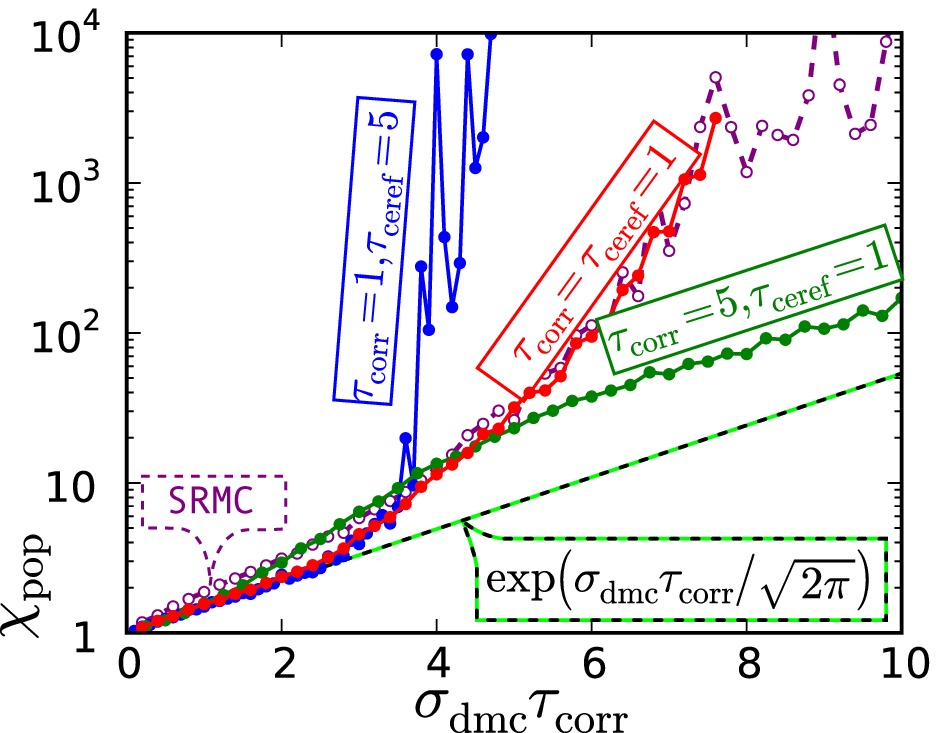}}}{(color
online) \label{fig:modeldmc}Inefficiency factor $\chi_{\operatorname{pop}}$ computed
for a model DMC process (see text) in dependence of the two relevant
parameters $\sigma_{\operatorname{dmc}} \tau_{\operatorname{corr}}$ and $\tau_{\operatorname{ceref}} /
\tau_{\operatorname{corr}}$. The exponential law extrapolated from the perturbative
limit is found to give a strict lower limit for $\chi_{\operatorname{pop}}$. Solid
circles refer to standard DMC with branching. Hollow circles refer to minimal
stochastic reconfiguration MC
(SRMC).\cite{buonaura-nsotthmuagfmctwafnow1998,assaraf-dmcmwafnow2000}}

\section{Hydrogen sample system}

To demonstrate our result in a real calculation, we have performed various DMC
runs using the CASINO program.\cite{casino2008} We chose a system of $N$
hydrogen atoms placed several thousand atomic units apart to make them
effectively independent. As a trial wave function, we used the exact ground
state with a detuning parameter $\alpha$ and an additional term to satisfy the
Kato cusp condition,\cite{kato-oteomsiqm1957} centered on each hydrogen
atom
\begin{eqnarray}
  \Psi_{\alpha} \left( r \right) & = & \alpha \mathrm{e}^{- \alpha r} + \left( 1 -
  \alpha \right) \mathrm{e}^{- \left( \alpha + 1 \right) r} .  \label{eqn:detune}
\end{eqnarray}
We performed a large variety of runs on this model system with system sizes $N
\in \left\{ 1, 2, 4, 8, \ldots, 64 \right\}$, and detuning parameters $\alpha
\in \left\{ 1.1, \ldots, 3.0 \right\}$ and target population $N_{\operatorname{pop}} =
200$. The DMC time step was set to $\tau_{\operatorname{step}} = 0.02$ in all cases.

The variance $\sigma_{\operatorname{dmc}}^2$ was found to be equal to
$\sigma_{\operatorname{vmc}}^2$ within the statistical error in all cases. In each
case, the population correlation factor $\chi_{\operatorname{pop}}$ was determined
from the variances using Eq.~(\ref{eqn:xipop-def}).

Obtaining a precise value for the correlation time $\tau_{\operatorname{corr}}$ takes
an extremely large amount of data in either of the two methods described
above. For a reasonably precise value, we performed a very long DMC run
($N_{\operatorname{step}} > 10^7$) on a single atom for each value of $\alpha$.
Several tests on larger systems confirmed that the same value holds for larger
numbers of atoms $N$. Since $\tau_{\operatorname{corr}}$ is independent of
$\tau_{\operatorname{step}}$ only if $\tau_{\operatorname{corr}} \gg \tau_{\operatorname{step}}$, we
performed these runs for the same step size as the main calculations.

As in Fig.~\ref{fig:modeldmc}, we plot the population correlation factor
$\chi_{\operatorname{pop}}$ from all our calculations as a function of the product
$\sigma_{\operatorname{dmc}} \tau_{\operatorname{corr}}$ and again find the exponential lower
bound described by Eq.~(\ref{eqn:xipop-exponential}), as displayed in
Fig.~\ref{fig:scaling}.

\tmfloatsmall{\resizebox{\columnwidth}{!}{\includegraphics{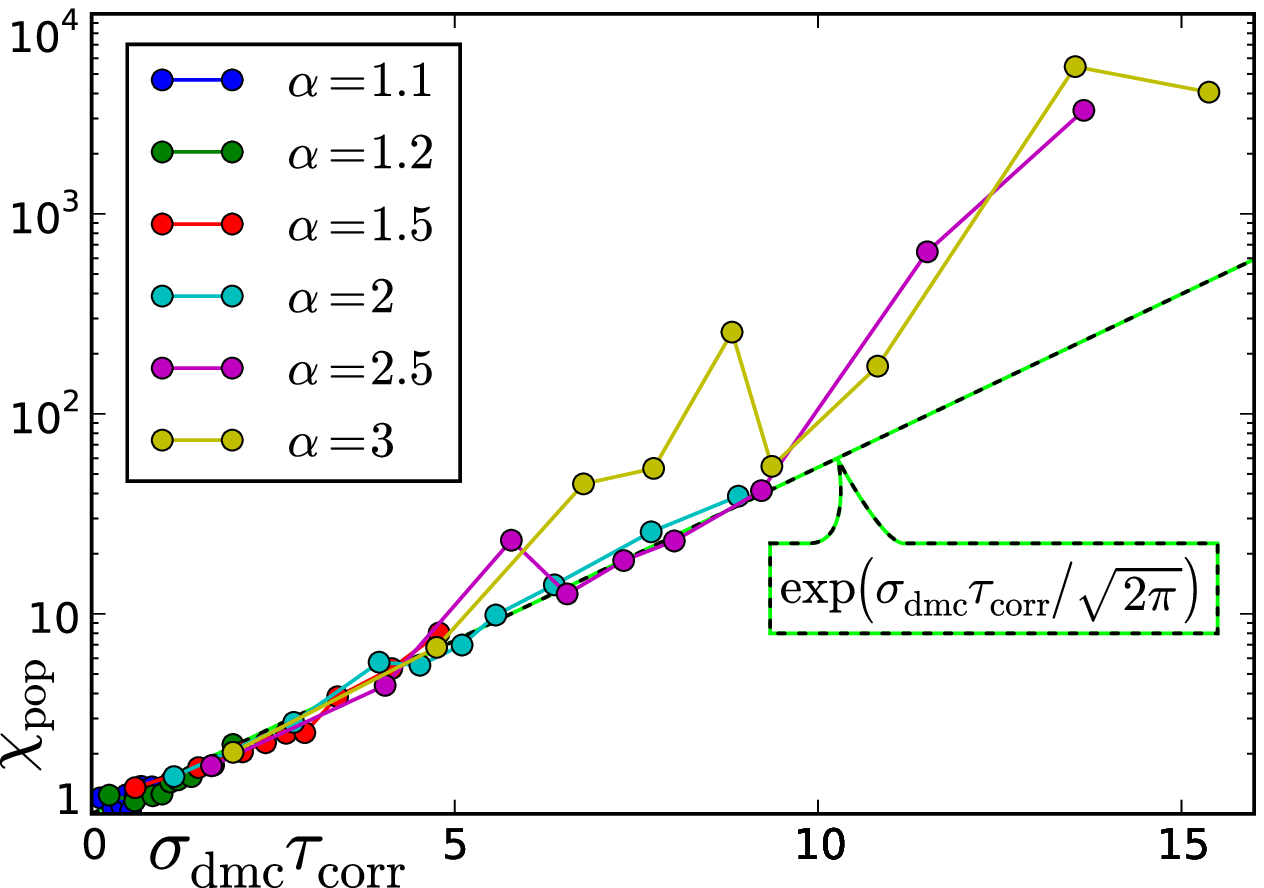}}}{(color
online) \label{fig:scaling}Data from many different calculations on a model
system of $N$ independent hydrogen atoms with a detunable trial wave function
[Eq.~(\ref{eqn:detune})]. $\sigma_{\operatorname{dmc}}$ and $\chi_{\operatorname{pop}}$ were
directly obtained from each run. $\tau_{\operatorname{corr}}$ was determined from a
single, very long run for each type of atomic wave function. Population
control is kept at the default $t_{\operatorname{ceref}} = 1$.}

\section{Analysis of various sample systems}

The exponential law in Eq.~(\ref{eqn:xipop-exponential}) has severe
implications for the scaling of the DMC method. Following
Eqs.~(\ref{eqn:vardmc}) and (\ref{eqn:ttotal2}), the total CPU cost becomes
\begin{eqnarray*}
  t_{\operatorname{total}} & \propto & \frac{\exp \left( X \sqrt{N} \right)}{\delta
  E_{\operatorname{atom}}^2} 
\end{eqnarray*}
or worse. So, even if population correlation may not be an issue yet for most
applications, it will eventually lead to an exponential scaling of the cost.
The factor $X$ can be reduced by optimizing the wave function, but the gain
that is possible with reasonable effort is very limited.

Table~\ref{tbl:samples} lists a selection of sample systems showing the size
at which the exponential scaling becomes observable. The integrated
correlation time $\tau_{\operatorname{corr}}$ [via Eq.~(\ref{eqn:tcorr})] and the raw
variance $\sigma_{\operatorname{dmc}}^2$ were computed for very small systems and
Eq.~(\ref{eqn:xipop-exponential}) was then used to estimate the size at which
a comparable system would show significant population correlation. All values
should be understood as rough estimates. The trial wave functions were either
taken from a library of examples or optimized with moderate effort. Further
optimizations could certainly reduce $\sigma_{\operatorname{dmc}}^2$ and thereby shift
the onset of significant population correlation. Typically, however,
significant effort is necessary even for minor improvements using
optimizations beyond the standard Jastrow terms.

\tmtablesmall{\begin{tabular}{llll}
  \textbf{atoms} (ae) & $\tau_{\operatorname{corr}}$ & $\sigma^2_{\operatorname{dmc}} /
  \operatorname{atm}$ & $\chi_{\operatorname{pop}} = 2$\\
  $\mathrm{\operatorname{He}}$ & 0.5 & 0.0044 & 2700 atoms\\
  $\mathrm{C}$ & 0.4 & 0.16 & 140 atoms\\
  $\mathrm{\operatorname{Ar}}$ & 0.04 & 8.0 & 250 atoms\\
  \textbf{molecules} (ae) &  & $\sigma^2_{\operatorname{dmc}} / \operatorname{mlc}$ & \\
  $\mathrm{H_2 O}$ & 0.1 & 0.58 & 550 molec.\\
  $\mathrm{\operatorname{CH}_4}$ & 0.3 & 0.24 & 120 molec.\\
  $\mathrm{C_2 H_4}$ & 0.4 & 0.51 & 38 molec.\\
  $\mathrm{\operatorname{SO}_2}$ & 0.06 & 7.5 & 105 molec.\\
  \textbf{crystals} &  & $\sigma^2_{\operatorname{dmc}} / \operatorname{atom}$ & \\
  diamond (pp) & 0.15 & 0.23 & 630 atoms\\
  diamond (ae) & 0.1 & 2.3 & 133 atoms\\
  graphite (pp) & 0.3 & 0.20 & 135 atoms\\
  silicon (pp) & 0.4 & 0.052 & 328 atoms\\
  \textbf{electron gas} &  & $\sigma^2_{\operatorname{dmc}} / \operatorname{elec} .$ & \\
  3d crystal ($r_s = 1$) & 0.2 & 0.26 & 193 elec.\\
  3d fluid ($r_s = 5$) & 5 & $4.2 \times 10^{- 4}$ & 330 elec.\\
  3d fluid ($r_s = 10$) & 16 & $5.1 \times 10^{- 5}$ & 242 elec.\\
  2d crystal ($r_s = 1$) & 0.4 & 0.038 & 570 elec.\\
  2d fluid ($r_s = 1$) & 0.3 & 0.033 & 1154 elec.
\end{tabular}}{\label{tbl:samples}Estimated values for various sample systems.
The last column gives the system size based on
Eq.~(\ref{eqn:xipop-exponential}) at which the population correlation becomes
significant with $\chi_{\operatorname{pop}} = 2$. Beyond this size, the DMC method
must be expected to become exponentially inefficient. The first three
categories are based on either all-electron (ae) or pseudopotential (pp) wave
functions with optimized Jastrow factors. All numbers should be understood as
rough estimates based on moderately optimized trial wave functions. Reducing
$\sigma_{\operatorname{dmc}}^2$ by further optimization will shift the onset of the
inefficiency by the same factor.}

\section{Alternative variants of QMC}

To this point the discussion was centered on the conventional DMC algorithm
including drift and branching. In the following, we will briefly discuss a
number of alternative QMC algorithms in view of the population correlation
scaling.

First, it is clear that population correlation can only be caused by some form
or branching. The variational MC (VMC) algorithm, which samples an explicitly
known wave function, clearly does not have this feature. A variant of DMC with
branching switched off (sometimes referred to as ``pure'' DMC) also features a
completely uncorrelated population. If all configurations are fixed to the
same statistical weight, this process produces the same distribution of
configurations and thereby the same total energy as VMC and can therefore be
seen as a variant of the former.

If, on the other hand, each configuration in a pure DMC run carries a
statistical weight evolving with the fluctuations in the local energy, the
effective population size $N_{\operatorname{pop}}^{\operatorname{eff}}$ is reduced in the same
way as it would be when branching were allowed, with the only difference that
decorrelation does not happen and the method becomes exponentially unstable
with simulation time.\cite{assaraf-dmcmwafnow2000}

A number of variants of the DMC algorithms keep the population size fixed and
include branching in form of stochastic reconfiguration, duplicating some
configurations and deleting
others.\cite{buonaura-nsotthmuagfmctwafnow1998,assaraf-dmcmwafnow2000,jones-andmcmadeoidoatsx2009}
While the population is fixed, the total weight is allowed to fluctuate
independently and the population control is replaced by a weight control
mechanism.

Our definition of the population correlation factor $\chi_{\operatorname{pop}}$ in
Eq.~(\ref{eqn:xipop-def}) is already kept general enough to capture the
effects of weight fluctuations within the population along the correlations
within the population and to capture the fluctuations of the total weight
along with the population fluctuations. Tests on several variants of the
branching strategy confirmed that these have no influence on the exponential
lower bound of $\chi_{\operatorname{pop}}$ but only affect how far the actually
measured $\chi_{\operatorname{pop}}$ exceeds the predicted exponential scaling (see
Fig.~\ref{fig:modeldmc}).

One remaining option to limit $\chi_{\operatorname{pop}}$ is the use of strong
population control on a small population, accepting a significant population
bias. The extreme case of this strategy would be a single walker with the
weight renormalized after every step, leading exactly to the VMC distribution.
We can, therefore, tune between exponentially scaling statistical inefficiency
and a population bias that ultimately leads to recovering the VMC algorithm,
which - as we know - does not suffer from exponential scaling.

Variants of QMC such as path integral MC (PIMC)\cite{ceperley-piittoch1995}
or reptation Monte Carlo (RMC)\cite{baroni-rqmcamfugaaic1999} are somewhat
related to DMC in the sense that they are based on a drift-diffusion process
in imaginary time. Unlike DMC, however, these methods are based on a true
Metropolis algorithm without the need for branching. Population correlation
does not occur and the statistical weight fluctuations are not a problem.
Instead, an analysis of the statistical efficiency of these methods would need
to take into account the correlation time and its scaling with system size.

\section{Conclusions}

To conclude, we have derived an expression for the scaling behavior of DMC
calculations when aiming at a fixed statistical precision per particle. Using
a linear scaling algorithm for an individual time step, constant scaling of
the total computational cost for the energy per particle is possible in
principle, except for a factor $\chi_{\operatorname{pop}}$, which quantifies the
correlation within the population of walkers. The exact value of
$\chi_{\operatorname{pop}}$ was derived in the perturbative limit, depending only on
the correlation time and the raw variance of the DMC process. Based on
numerical evidence, we demonstrated that an exponential extrapolation of the
perturbative law gives a strict lower bound to the inefficiency factor
$\chi_{\operatorname{pop}}$. From this, it follows that the DMC algorithm generally
scales at least exponentially in the square root of the system size.

The numbers for actual sample systems indicate that this exponential scaling
should not even be observable in most DMC based studies done so far, leaving
plenty of room to do interesting research with the DMC method.

Alternative schemes for branching and population control that have been
suggested\cite{buonaura-nsotthmuagfmctwafnow1998,assaraf-dmcmwafnow2000,jones-andmcmadeoidoatsx2009}
may certainly influence the efficiency of the algorithm. The exponential lower
bound of the statistical inefficiency, however, may at best be shifted over
towards an exponentially scaling population control
bias.\cite{drummond-feicqmcc2008}

It must be stressed that this exponential scaling factor is specific to the
DMC method and does not occur in other methods like VMC. It is not linked to
the more fundamental fermion sign problem\cite{troyer-ccafltfqmcs2005} and
it is not limited to certain observables.\cite{warren-psbiddqmcs2006}

In fact, the exponential scaling may be a symptom of the very nature of the
DMC process. In general, Markov-chain MC methods such as VMC exhibit excellent
scaling behavior. DMC however, is not based on a Markov process but rather on
the simulation of a time dependent stochastic diffusion process. As such, it
must be expected to suffer from the exponential accumulation of errors
inherent in the simulation of time evolution in non-integrable systems. The
population control that is necessary to stabilize the process might then
necessarily lead to exponential scaling either in the bias or the efficiency
of the process. The only way to overcome this problem might then be to resort
to alternative QMC methods like VMC, PIMC or RMC that are based the stochastic
computation of a multi-dimensional integral in the original spirit of
Markov-chain MC methods.

\section{Acknowledgments}

We acknowledge fruitful discussions with Richard Needs, Neil Drummond and
Matthew Foulkes. This work was funded by the DAAD and the EPSRC. The
computations were performed using the facilities of the University of
Cambridge High Performance Computing Service.

\end{document}